\documentclass{conm-p-l}

\copyrightinfo{2018}{}

\usepackage{graphicx}
\usepackage{subfigure}

\usepackage{amssymb,amsmath,amsthm,amscd}

\theoremstyle{definition}

\theoremstyle{remark}

\numberwithin{equation}{section}

\newcommand{\e}{\epsilon}

\newcommand{\Dl}{\Delta}

\newcommand{\ra}{\rightarrow}
\newcommand{\al}{\alpha}

\newcommand{\sg}{\sigma}

\newcommand{\pa}{\partial}

\newcommand{\na}{\nabla}

\begin{document}

\title[Dependence upon initial conditions]{Dependence upon initial conditions}

\author{Y. Charles Li}
\address{Y. Charles Li, Department of Mathematics, University of Missouri, 
Columbia, MO 65211, USA}
\email{liyan@missouri.edu}
\urladdr{http://faculty.missouri.edu/~liyan}

\curraddr{}
\thanks{}

%\subjclass{Primary 76, 35; Secondary 34}


\date{}

\dedicatory{}


\begin{abstract}
This article discusses dependence on initial conditions in natural and social sciences with 
focus on physical science. The main focus is on the newly discovered rough dependence on initial data.
\end{abstract}

\maketitle

\section{Introduction}

        The evolution of a natural phenomenon starts from and depends upon its initial condition. The dependence upon the initial condition is an extremely complicated matter. For some phenomena like harmonic oscillations, the dependence upon the initial condition is simple. But the simplicity of the dependence can easily be broken, for instance, when the harmonic oscillations are forced and damped, the dependence upon the initial condition can be sensitive (chaotic), leading to long term unpredictability. Now chaotic phenomena have been well recognized. Chaotic phenomena appear in both simple and complicated systems such as the Lorenz system and the Rayleigh-B\'enard convection. On the other hand, for complicated systems like fluid turbulent flows, the dependence upon the initial condition is often rough, leading to short term unpredictability \cite{FL18}. Short term unpredictability is very close to total randomness. Even though they are deterministic, the governing equations of turbulence, i.e. the Navier-Stokes equations have little control over the evolution of turbulence. For random phenomena like Brownian motions, the dependence upon the initial condition is weak. So far, we have mainly discussed phenomena in physical sciences, governed by the initial value problems of differential equations. Such phenomena can be described with the tools of calculus. 
These physical sciences include physics, space science and earth science. Some chemical reactions can also be described via differential equations. 

        For natural phenomena in biological sciences, differential equations and calculus tools 
do not offer successful descriptions. Statistical tools are more successful. Even though they 
start from and depend upon initial conditions, biological evolutions are not 
successfully described by differential 
equations and calculus tools due to the lack of laws such as Newton's laws or other physical laws. Biological agents have intelligence, thus they do not follow the laws like those in physics (e.g. law of gravitation) for non-intelligent objects.
Thus the dependence upon the initial condition for biological phenomena is more subtle than that for 
physical phenomena, and has an intelligence factor in it. The intelligence factor is hard to describe 
via calculus tools. Actually statistical tools seem more suitable to describe the intelligence factor. 
There could be good mathematical tools to be discovered to accurately describe the intelligence factor.

        For social phenomena, the intelligence factor is even stronger. In fact, social phenomena are 
difficult to quantify, for example, how to quantify a thought? But we can still talk about evolution 
of social phenomena and their dependence upon initial conditions. Nowadays there are interests in 
studying interdisciplinary subjects between social science and biological science, like collective 
behaviors of biological agents - fish schooling, bird flocking, amoebae fruiting etc. Collective behavior of humans is a subject of social science. Intelligence level of agents can be a measure that distinguish among social, biological and physical phenomena. Social agents i.e. humans have the highest intelligence level, other biological agents have lower intelligence levels, and physical agents have no intelligence.

\section{Solution operator and solution's derivative operator}

The evolution of a physical phenomenon is often described (governed) by an initial value problem of differential equations
\begin{equation}
\pa_t u = F(u, \pa_x) , \quad u(0,x) = u_0(x) , \label{IVP} 
\end{equation}
where $t$ is the time variable, $x$ is the space variable, and $u$ is the description variable of the 
phenomenon. When there is no space variable, the differential equations are ordinary differential equations, otherwise partial differential equations. Denote by $S^t$ the solution operator which 
maps the initial value to the value of solution at time $t$, and $t$ parameterizes the solution operator. 
\[
S^t (u_0) = u(t) .
\]
We can view the initial value as a point in a 
space - called phase space. Thus for any fixed $t$, the solution operator $S^t$ is a map defined in the phase space. For ordinary differential equations, the phase space is the Euclidean space. For partial differential equations, the phase space is a function space e.g. a Sobolev space. A solution can be viewed as an orbit in the phase space. For ordinary differential equations, the orbit can be differentiable (i.e. $S^t$ can be differentiable in $t$). For partial differential equations, the orbit is usually only continuous. The solution's property in the space variable is reflected by the nature of the phase space. The 
basic question on a solution is its existence and uniqueness. But the patterns of all orbits 
in the phase space are more characteristic of the 
physical phenomena. The pattern of all orbits is parametrized by the initial conditions, and characterized by the solution's dependence upon initial conditions. The basic Hadamard requirement
for well-posedness is that the solutions depend on their initial conditions continously. But it 
turns out that the differential property of solution's dependence upon initial conditions is more crucial in characterizing the pattern of all orbits. The differential property characterizes how
the solutions react to perturbations, i.e. how the nearby orbits are fibered in the phase space. 
\[
du(t) = \na_{u_0} S^t (u_0)  \circ d u_0 ,
\]
where $d u_0$ is the initial differential which can be viewed as an infinitesimal perturbation. The 
temporal property of the solution's derivative operator 
\[
 \na_{u_0} S^t (u_0), 
\]
describes how the initial perturbation is evolved. Since the time dependence of the solution's 
derivative operator can be wild, the patterns of all orbits can be wild too, reflecting the wild nature of physical phenomena ranging from chaos to turbulence. We can define two types of orbits:
\begin{itemize}
\item {\it A Mathematical Orbit} is given by the unique solution to the initial value problem 
(\ref{IVP}). 
\item {\it A Physical Orbit} is a realization of the evolution of the physical phenomenon. It 
incorporates all the perturbations and corrects the inaccuracy of the mathematical model (\ref{IVP}).
\end{itemize}
Sometimes, a mathematical orbit can be an accurate approximate to a physical orbit, e.g. the 
simple harmonic oscillation. But a mathematical orbit can often be fundamentally different 
from a physical 
orbit. A physical orbit always carries perturbations which can dramatically drift the physical orbit
away from the mathematical orbit. The ever existing perturbations can appear all the time along a physical orbit, and each such a perturbation can be amplified by the solution's derivative operator. 
One can view all the mathematical orbits as forming a fiber net in the phase space, and a physical orbit 
drifts along the net by perturbations. When it comes to perturbations, the most important thing is whether they grow or decay. If they decay, then the perturbations are of little significance. This 
scenario will be reflected by the mathematical orbit net being converging in that region. If they grow, then the perturbations can significantly drift the physical orbits away from the mathematical 
ones, and the mathematical orbit net in that region is diverging. Of the greatest interest is the 
fastest growth of the perturbations, which is given by the norm of the solution's derivative 
operator
\begin{equation}
\| \na_{u_0} S^t (u_0) \| = \sup_{d u_0} \frac{ \| du(t) \|}{\| du_0 \|}. \label{NSD}
\end{equation}
Notice that at different time $t$, the fastest growing initial perturbation $du_0$ may be different.
Generically all perturbations can be present, and the fastest growing perturbation will quickly 
prevail.

\section{Sensitive dependence upon initial conditions - long term unpredictability}

When the norm of the solution's derivative operator grows exponentially in time, 
\[
\| \na_{u_0} S^t (u_0) \| \sim e^{\sg t} , \quad \sg > 0, 
\]
some perturbations amplify exponentially in time too. When the time is long enough, such 
perturbations are amplified to a substantial magnitude such that nonlinear effect is significant. 
If there is a region in the phase space, that traps the orbits, and for a lot of initial 
conditions $u_0$, the norm of the solution's derivative operator grows exponentially in time, 
the physical orbit must be chaotic! First of all, the physical orbit is trapped in the region, 
second the ever-existing perturbations kept being amplified exponentially along the physical orbit. 
The exponential amplifying of perturbations reveal the exponential diverging nature of the mathematical solution net. In fact, it is not just the physical orbit being chaotic, the mathematical 
orbit is chaotic too, even though there is no perturbation. Some of the mathematical orbits 
in chaos are in one-to-one correspondence with the binary Bernoulli shift orbits which are chaotic. 
This has been established both in ordinary differential equations \cite{Pal84} and in partial differential equations \cite{Li03} \cite{Li04}. In ideal situations, chaotic physical orbits have shadowing mathematical orbits nearby. Thus the underlying mathematical orbit net in the phase space can be chaotic too. We will talk more on this in details later on. 

In the chaos situation, the solution's dependence upon initial conditions is called ``sensitive".
No matter how small its initial perturbation is, the exponentially amplified perturbation will reach 
a substantial magnitude when the time is long enough. This is the so-called long 
term unpredictability. 

It is natural to expect that sometimes the norm of the solution's derivative operator grows exponentially in time, since when $u_0$ is a fixed point, the unstable eigenvalues will lead to 
exponential growths. 

In more general situations, the exponential growth of the norm of the solution's derivative 
operator is usually pursued by calculating the Liapunov exponents
\[
\sg = \lim_{t \ra \infty} \frac{1}{t} \text{ ln }  \frac{ \| du(t) \|}{\| du_0 \|}
    = \lim_{t \ra \infty} \frac{1}{t} \text{ ln }  \frac{ \| \na_{u_0} S^t (u_0)  
\circ d u_0\|}{\| du_0 \|}.
\]
There can be a spectrum of Liapunov exponents given by different initial differential basic directions. 
Under ideal situations, the spectrum of Liapunov exponents can be independent of the initial 
condition $u_0$, as shown by Oseledets theorem \cite{Ose68} \cite{Rue79} \cite{Man83}. 
Numerical calculations 
usually reveal the largest Liapunov exponent which prevails in long time starting from a 
generic initial perturbation $du_0$ which contains components from all the initial differential basic
directions. Numerical verification of a positive Liapunov exponent in an orbit trapping region is usually a good indicator of a chaotic dynamics.

\section{Rough dependence upon initial conditions - short term unpredictability}

Of course, the norm of the solution's derivative operator can grow in time in various ways. 
Exponential growth is only one of them. For Navier-Stokes equations of fluids,
\[
\pa_t u = - u \cdot \na u - \na p + \frac{1}{Re} \Dl u,  \quad \na \cdot u = 0, 
\]
we have an upper bound on the norm of the solution's derivative operator \cite{Li14},
\begin{equation}
\| \na_{u_0} S^t (u_0) \| \leq e^{\sg \sqrt{Re} \sqrt{t} + \sg_1 t}, \label{UB}
\end{equation}
where $\sg_1 = \sqrt{\frac{e}{2}} \sg$ only depends on the base solution starting 
from the initial condition $u_0$.  In view of the fact that for Euler equations of 
inviscid fluids, 
\[
\pa_t u = - u \cdot \na u - \na p ,  \quad \na \cdot u = 0, 
\]
the solution's derivative never exists \cite{Inc12} \cite{Inc15}, we proposed the following 
\begin{itemize}
\item {\bf Conjecture:} The upper bound (\ref{UB}) is always realized at any initial 
condition in fully developed turbulence.
\end{itemize} 
The conjecture is stronger than saying that the upper bound is sharp, which only requires one 
initial condition $u_0$ realizing the upper bound. Numerical simulations \cite{FL18} verified 
that initial stage amplification of perturbations at large Reynolds number indeed follows 
\[
e^{ c \sqrt{t}}.
\]
We believe that the $\sqrt{Re}$ nature in the exponent of the upper bound (\ref{UB}) can only be 
revealed after taking the supremum over all initial perturbations $du_0$ as in (\ref{NSD}). This will 
be a difficult task for numerical simulations. Analytically proving the above conjecture is 
going to be a daunting task. 

Beyond the above conjecture, we believe that the upper bound (\ref{UB}) are often realized even 
when the Reynolds number is not large. Based upon the realization of the upper bound (\ref{UB}),
when the time is small, the first term in the exponent dominates, and this term can cause the amplification to be superfast when the Reynolds number is large. Even during a short time, the 
perturbation can be amplified to a substantial size. We call this ``short term unpredictability". Notice also that the derivative of $\sqrt{t}$ at $t=0^+$ is infinity. By the 
time $t \sim Re$, the two terms in the exponent are about equal. After the time $t \sim Re$, the second term dominates, and this term is the classical Liapunov exponent that causes chaos (long term unpredictability). Thus the time $t \sim Re$ is the temporal separation point between short term unpredictability and long term unpredictability. When the Reynolds number is large, long before the separation point $t \sim Re$, the first term in the exponent already amplifies the perturbation to substantial size (such that nonlinearity is crucial), and the second term does not get a chance to dominate. Thus fully developed turbulence is dominated by such short term unpredictability. When the
Reynolds number is moderate, both terms in the exponent have a chance to dominate, and the corresponding (often) transient turbulence is dominated by chaos in long term.

When the Reynolds number approaches infinity, the upper bound (\ref{UB}) approaches infinity, and this is in consistent with the fact that the solution's derivative never exists for Euler equations of inviscid fluids \cite{Inc12} \cite{Inc15}. Explicit examples were also constructed to show that
the norm of the solution's derivative is infinity for Euler equations, and approaches infinity when 
the Reynolds number approaches infinity for Navier-Stokes equations \cite{Li17}. 

The short term unpredictability is very different from the long term unpredictability of chaos. 
The short term unpredictability is closer to total randomness than the long term unpredictability of chaos. Nevertheless, the short term unpredictability is still not total randomness, for instance the solution operators of Euler equations and Navier-Stokes equations are still continuous in their initial data. There are also conserved quantities under the Euler dynamics, that do not change 
superfast under perturbations. Such a short term unpredictability leads to a peculiar process that is very close to a random process but still constrained. When the Reynolds number is moderate, dynamics of Navier-Stokes equations is quite far away from that of Euler equations. The norm of the solution's derivative operator is moderate. Turbulence at such a stage is often transient, and bears clear resemblance to finite dimensional chaos. One can name such turbulence as chaos in Navier-Stokes equations. 

We believe that the short term unpredictability causes the abrupt nature in the development of high Reynolds number turbulence. Since perturbations constantly exist, there are constantly such superfast amplifications of perturbations which lead to the persistence nature of high Reynolds number turbulence (so-called fully developed turbulence) in contrast to the transient nature of moderate Reynolds number turbulence.

In terms of phase space dynamics, when the Reynolds number is very high, fully developed turbulence is not the result of a strange attractor, rather a result of superfast amplifications of ever present perturbations - short term unpredictability. Strange attractor is a long time object, while the development of such violent turbulence is of short time. Such fully developed turbulence is maintained by constantly superfast perturbation amplifications. When the Reynolds number is set to infinity, the perturbation amplification rate is infinity. So the dynamics of Euler equations is very close to a random process. In contrast, chaos in finite dimensional 
conservative systems often manifests itself as the so-called stochastic layers. Dynamics inside the stochastic layers has the long term sensitive dependence on initial data.

For chaotic phenomena, the mathematical model (\ref{IVP}) is not an accurate model in 
long term, but is a good model in short term. On the other hand, for fully developed turbulence,
the mathematical model, i.e. the Navier-Stokes equations are not a good model even in short term. 

The Liapunov exponent is a measure of long term unpredictability. When the norm of the solution's derivative operator can be large in short time, short term unpredictability results. One can define the following exponent to measure short term unpredictability
\[
\eta = \lim_{t \ra 0^+} \frac{1}{t^\al} \text{ ln } \|  \na_{u_0} S^t (u_0) \|, \quad \text{where } 
\al > 0 .
\]
When $\eta$ is large (e.g. approaching infinity as a parameter approaches a limit), one has short 
term unpredictability. In the case of Navier-Stokes equations, $\eta$ can be as large as
\[
\eta = \sg \sqrt{Re},
\]
with $\al = 1/2$, where $\sg$ is given in (\ref{UB}).

\section{Other dependence upon initial conditions}

There may be other ways yet to be discovered, in which the norm of the solution's derivative operator can grow in time. In general, there are two time limits: long term limit and short term limit. For short 
term limit,
\[
\| \na_{u_0} S^t (u_0) \| \ra 1, \quad \text{as } t \ra 0^+ .
\]
Even for the exponential type growth,
\[
\| \na_{u_0} S^t (u_0) \| \sim e^{f(t)},
\]
$f(t)$ can take other forms yet to be discovered. 

Physical phenomena constantly suffer from the ever existing perturbations. The mathematical 
model (\ref{IVP}) of a physical phenomenon usually does not take into account the ever 
existing perturbations in the physical phenomenon. When the perturbations are amplified in 
the physical phenomenon, the mathematical model will reflect the amplification via the growing 
norm of the solution's derivative operator. Thus we believe in the doctrine that physical 
phenomena such as chaos and turbulence, driven by perturbations are reflected in the 
structure of the mathematical models. 

Due to the intelligence factor, one initial condition may lead to multiple evolutionary orbits 
in biological and social phenomena, for instance, starting from the same time flash of a bird 
flock, repeating the future flock flying may lead to very different flock orbits. To the entire bird
flock, the intelligence factor of individual birds seems a random factor, but the intelligence factor
keeps the collective behavior of the flock. The intelligence factor is a stability factor which 
usually suppresses the effect of the small random perturbations.

\section{Shadowing of pseudo-orbits}

Physical orbits and mathematical orbits are clearly different. But do they have any connection? 
It turns out that, in ideal situations of chaotic dynamics, mathematical orbits can shadow physical orbits. The simplest way to illustrate the shadowing property is through a discrete time solution 
operator $S^n$, where $n$ is an integer variable. A pseudo-orbit is a model for a physical 
orbit. A pseudo-orbit is such a sequence 
\[
u_n , \quad  n = 0, \pm 1, \pm 2, \cdots 
\]
that 
\[
\| S^1(u_n) - u_{n+1} \| < \e ,  \quad  n = 0, \pm 1, \pm 2, \cdots 
\]
where $\e$ is small. That is, at every time step, the pseudo-orbit may have a small perturbation. 
In ideal situations of chaotic dynamics, e.g. Anosov system \cite{Ano67} \cite{Sin72}, Axiom A 
system \cite{Bow75}, and continuous transversal homoclinic dynamics \cite{Pal84} \cite{Li03} \cite{Li04}, every pseudo-orbit has a shadowing mathematical orbit which is nearby for all time. 
By choosing two mathematical orbit-segments such that one segment's end is close to the beginning
of the other \cite{Pal84} \cite{Li03}, and labeling the two segments by $0$ and $1$, one can 
associate every binary sequence with a pseudo-orbit which in turn is shadowed by a mathematical 
orbit, and the action of the solution operator on the mathematical orbit is corresponding to 
the Bernoulli shift on the binary sequence. Bernoulli shift can be regarded as a prototype for 
chaotic dynamics. Thus, the mathematical orbits are chaotic too, and they shadow the chaotic 
physical orbits. From the shadowing perspective, mathematical orbits and mathematical 
models in general are still significant in modeling chaotic phenomena. 

The natural and important question is: 
\begin{itemize}
\item Does rough dependence dynamics have a shadowing property?
\end{itemize}
An answer to this question will answer from the shadowing perspective the following more general 
questions: In the rough dependence situations,
\begin{itemize}
\item are mathematical orbits relevant to the physical phenomena?
\item are the mathematical models significant to the physical phenomena?
\end{itemize}
From the initial value problem perspective, rough dependence leads to short term unpredictability,
and the mathematical and physical orbits starting from the same initial condition are very different 
even in short time. Thus mathematical orbits seem not relevant to the physical phenomena, and 
the mathematical models are not significant to the physical phenomena anymore.

\section{Dependence upon parameters and bifurcations}

When the differential equation depends on a parameter $\nu$, the stable state (attractor) changes 
with the parameter, this is the so-called bifurcation. 
\begin{equation}
\pa_t u = F(u, \pa_x, \nu ) , \quad u(0,x) = u_0(x) , \label{BIV} 
\end{equation}
The parameter $\nu$ can be incorporated into the initial condition of an enlarged equation,
\begin{equation}
\pa_t u = F(u, \pa_x, \nu ) , \ \pa_t \nu = 0, \quad u(0,x) = u_0(x) ,\  \nu (0) = \nu . \label{EBI} 
\end{equation}
The dependence of the solution operator to the enlarged equation
\[
S^t(u_0, \nu )
\]
on $\nu$ in long term is the bifurcation problem. For fixed $u_0$, changing $\nu$, the asymptotic
limit of $S^t(u_0, \nu )$ as $t \ra \infty$ can be quite different. The asymptotic limits (attractors) are usually not just a point, rather a more general set. The value of $\nu$ at which the attractor changes is called the bifurcation point.

\end{document}